# Spin-valley-locked Electroluminescence for High-Performance Circularly-Polarized Organic Light-Emitting Diodes

Yibo Deng,[1,†] Teng Long,[1, †] Pingyang Wang,[1] Han Huang,[1] Zijian Deng,[1] Chunling Gu,[2,*] Cunbin An,[1] Bo Liao,[3] Guillaume Malpuech,[4] Dmitry Solnyshkov,[4,5*] Hongbing Fu,[1,3,*] Qing Liao[1,*]

[1]Beijing Key Laboratory for Optical Materials and Photonic Devices, Department of Chemistry, Capital Normal University, Beijing 100048, China

[2]Institute of Process Engineering, Chinese Academy of Sciences, Beijing, 100190, China

[3]School of Materials Science and Engineering, Hunan University of Science and Technology, Xiangtan, Hunan 411201, China

[4]Université Clermont Auvergne, Clermont Auvergne INP, CNRS, Institut Pascal, F-63000 Clermont-Ferrand, France

[5]Institut Universitaire de France (IUF), 75231 Paris, France

†These authors contributed equally: Yibo Deng, Teng Long.

**Abstract**

Circularly polarized (CP) organic light-emitting diodes (OLEDs) have attracted attention in potential applications including novel display and photonic technologies. However, conventional approaches cannot meet the requirements of device performance, such as high dissymmetry factor, high directionality, narrowband emission, simplified device structure and low costs. Here, we demonstrate spin-valley-locked CP-OLEDs without chiral emitters, but based on photonic spin-orbit coupling, where photons with opposite CP characteristics are emitted from different optical valleys. These spin-valley locked OLEDs exhibit a narrowband emission of 16 nm, a high EQE of 3.65, a maximum luminance of near 98000 cd/m$^2$ and a $g_{EL}$ of up to 1.80, which are among the best performances of active single-crystal CP-OLEDs, achieved with a simple device structure. This strategy opens an avenue for practical applications towards three-dimensional displays and on-chip CP-OLEDs.

# Introduction

Organic circularly polarized (CP) electroluminescence (EL) devices with narrowband emission have attracted extensive attention due to their potential applications in three-dimensional (3D) displays, quantum computing, and other photonic technologies[1-4]. Typically, CP light can be generated from non-polarized EL by passing it through a linear polarizer and a quarter-wave plate as in commercial 3D display panels[5]. Beyond its bulky character, this solution always leads to a marked loss in brightness and decrease in contrast. A prospective strategy is the active generation of CP-EL directly from organic light-emitting diodes (OLEDs)[6-8]. Great efforts have been devoted to developing new CP-EL materials, such as the emission layer of OLEDs based on the chiral perturbation strategy[9,10] or chiral assemblies of achiral molecules owing to symmetry breaking[11,12], and revealed the relationship between the chiral structures and CP properties[13,14]. These approaches often involve complicated synthesis and stringent purification methods, however, the performances of CP-OLEDs, such as dissymmetry factor ($g_{EL}$), color purity and luminance, are not always satisfactory[15]. Importantly, to build upon the significant recent progress of virtual reality (VR)/augmented reality (AR) technologies[16], for example, OLEDs with up to 50000 cd/m$^2$ in Apple Vision Pro, it is extremely desirable to produce CP-OLEDs with high luminance, narrowband, high $g_{EL}$, and chiral-emitter free.

Achiral organic materials present a promising prospective for OLEDs due to the excellent device performances with external quantum efficiency (EQE) and tunable bandgap[10,17,18]. However, they are not suitable for producing CP-EL due to the absence of an intrinsic spin-locking mechanism in electronic structures of these materials[7,8]. Similar to the spin degree of freedom of electrons in condensed matter physics, the polarization degree of freedom of photons can also be manipulated through photonic spin-orbit coupling (SOC), where the polarization of photons is akin to the spin of electrons[19,20]. The presence of photonic SOC related to cavity-induced transverse-electric/transverse-magnetic (TE-TM) splitting and emergent optical activity (also called Rashba-Dresselhaus (RD) SOC) has been demonstrated in micro- and nano-scale structures, including two-dimensional (2D) inorganic materials, liquid crystals, and organic materials, and has led to significant advances in topological photonics[21-24]. Among them, organic crystals have developed rapidly because their large anisotropy, combining both types of photonic SOC, provides rich synthetic effective Hamiltonians to manipulate the polarizations of cavity photons and realize a nontrivial band geometry of optical bands[23,25,26]. This advantage of organic microcrystals with a pronounced photonic SOC provides new opportunities to develop CP-OLEDs using achiral materials. Recently, some of us achieved CP-OLEDs with high $g_{EL}$ in achiral device architectures by means of RD spin-orbit coupling[27]. However, their unsatisfactory EQE and luminance, which might be attributed to the unbalanced charge injection and transport[28,29], hinders their practical application. Additionally, the CP-EL beams of opposite polarization were generated close to the normal direction, leading to spatial mixing in the far field and decreasing the contrast.

Here, we construct synthetic tilted Dirac valleys in the photonic band of OLED architectures and endow the spin-valley-locking feature to the CP-EL with a narrowband emission. Organic single microcrystals of (2Z,2'Z)-3,3-([1,1'-biphenyl]-4,4'diyl)bis(2-(naphthalen-2-yl)acrylonitrile (BPDBNA), possessing balanced charge injection and transport properties, are designed and synthesized by a physical vapor deposition method. We construct the OLED devices, where BPDBNA microcrystals sandwiched between two silver layers and molybdenum trioxide ($MoO_3$)/$C_{60}$ and 1,3,5-Tris(1-phenyl-1H-benzimidazol-2-yl)benzene (TPBi) were chosen for hole and electron transport layers, respectively. Based on this simple device configuration, we obtain the chiral-emitter free CP-OLEDs with an EQE of 3.65, a maximum luminance of near 98000 cd/m$^2$ and a $g_{EL}$ of up to 1.80, which are among the best performances of active single-crystal CP-OLEDs. The electroluminescence with opposite CP characteristics is selectively associated with two optical valleys, whose emission angles are respectively 20º and -20º. These unique spin-valley-locked CP-OLEDs with top-emitting architecture demonstrate a promising strategy for light sources, display technologies and spin photonic applications.

## Results and Discussion

**Principle of spin-valley-locked emission from organic microcavities.**

In planar microcavities filled with organic anisotropic crystals, it was demonstrated that the combined photonic SOCs gives rise to a synthetic effective field acting on cavity photons (as sketched in Figure 1a), accounting for the polarization effects arising in this structure[23]. In this case, two orthogonally linearly polarized cavity modes of opposite parity meet at two special points along the $k_x$ axis because of the large anisotropy of organic crystals. Everywhere else the degeneracy is lifted by the TE-TM splitting. At the two degeneracy points, the splitting between the circular-polarized modes of opposite chirality is induced by the emergent optical activity (also called RD SOC with equal strength), recently discovered and explained in highly-birefringent cavities with liquid crystals[30] and organic materials[23]. Figure 1b shows two simulated Hamiltonian eigenmodes of opposite parity. The energy dispersions of the two bands anti-cross at two discrete points along $k_x$ in the 2D momentum space. Close to the anticrossing points, these bands manifest a valley-dependent Berry curvature and nontrivial band geometry (Figure 1c), in which each tilted gapped Dirac cone of the valleys carries a topological charge $\pm 1/2$[19,20]. In such organic microcavities, photons are selectively emitted from opposite photonic valleys with opposite spins (σ+ and σ−) under optical pumping[26]. Such spin-valley locking of emitted photons can be considered as a consequence of non-zero Berry curvature ($B_z$) for a given band, which one can measure as $B_z = \frac{1}{2}\sin\theta(\partial k_x \theta \partial k_y \phi - \partial k_y \theta \partial k_x \phi)$, where $\theta(k)$ and $\phi(k)$ are the polar angle and the azimuthal angle of the Stokes vector on the Poincaré sphere[23]. Hence, $B_z$ can be extracted from the measured three Stokes vector S($k$) components of the modes, which are determined by the six different light polarizations (horizontal-vertical, diagonal-anti-diagonal, and left-right circular).

The Stokes vector can be found theoretically from the eigenvectors of an effective Hamiltonian $H(\mathbf{k}) = H_{\text{TE-TM}} + H_{\text{XY}} + H_{\text{OA}}$, that we write in the circular polarization basis, where $H_{\text{TE-TM}} = \beta_1(k_x^2 - k_y^2)\hat{\sigma}_x + 2\beta_1 k_x k_y \hat{\sigma}_y$ describes the intrinsic TE-TM splitting of cavity modes, $H_{\text{XY}} = (\beta_0 + \beta_2 \mathbf{k}^2)\hat{\sigma}_x$ is the Hamiltonian representing the

XY splitting where the $k^2$ term is due to the mass difference for horizontally and vertically-polarized modes with different mode numbers, $H_{OA} = 2\alpha\hat{\sigma}_z k_x$ is the emergent OA Hamiltonian. The above effective Hamiltonian can be written in the form of a 2 × 2 matrix:

$$H = \begin{pmatrix} E_0 + \frac{\hbar^2}{2m}k^2 - 2\alpha k_x & \beta_0 + \beta_1 k^2 e^{-2i\varphi} + \beta_2 k^2 \\ \beta_0 + \beta_1 k^2 e^{2i\varphi} + \beta_2 k^2 & E_0 + \frac{\hbar^2}{2m}k^2 + 2\alpha k_x \end{pmatrix} \quad (1)$$

where $E_0$ is the energy of the ground state, $m$ is the effective mass of cavity photons, $\beta_1$ is the strength of the TE-TM splitting, $\beta_2$ characterizes the difference of H and V effective masses. $\beta_0 = E_X - E_Y$, $E_X$ and $E_Y$ are the ground state energies of X-polarized and Y-polarized modes of opposite parity here. $\alpha$ describes the emergent OA, and $\varphi$ ($\varphi \in [0, 2\pi]$) is the polar angle of the wave vector $k$.

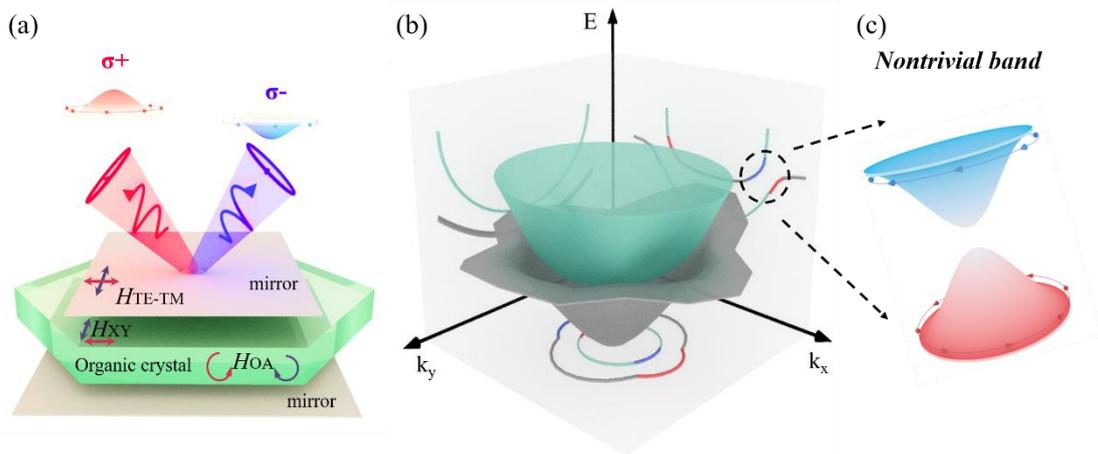

**Figure 1. Schematics of organic microcavity and spin-valley-locking emission.** (a) Schematic diagram of an optical cavity consisting of two mirrors and filled with birefringent microcrystals. (b) Three-dimensional polarization-resolved inverse spatial tomography imaging. (c) The nontrivial band geometry.

**BPDBNA molecular design and single-crystal OLED.**

To experimentally realize electrically-driven spin-valley-locking CP-OLEDs, we first design a novel organic molecule BPDBNA for the improvement of device performances. Currently, one of the main reasons for limiting device EQEs is the high electron affinity and unbalanced charge injection and transport[28,29,31]. The effective strategy to overcome some of these limitations is to adopt an organic material with an electron affinity of ~ 3.6 eV (or higher), which reduces the non-radiative recombination losses and enhances the air stability of the device due to the use of non-reactive electron-injection layers[3,31]. Theoretical simulations indicate that cyano-group (-CN) can effectively reduce the lowest unoccupied molecular orbital (LUMO) level of organic molecules to the electron affinity of ~3.6 eV (Figure S1). Therefore, we designed BPDBNA molecule (Figure 2a), in which a bi(phenyl-vinyl) was chosen as the backbone and bridged two naphthalene rings at its two ends, ensuring excellent charge transport and efficient emission. Then two -CN were introduced on the positions of the double bond to decrease the electron affinity.

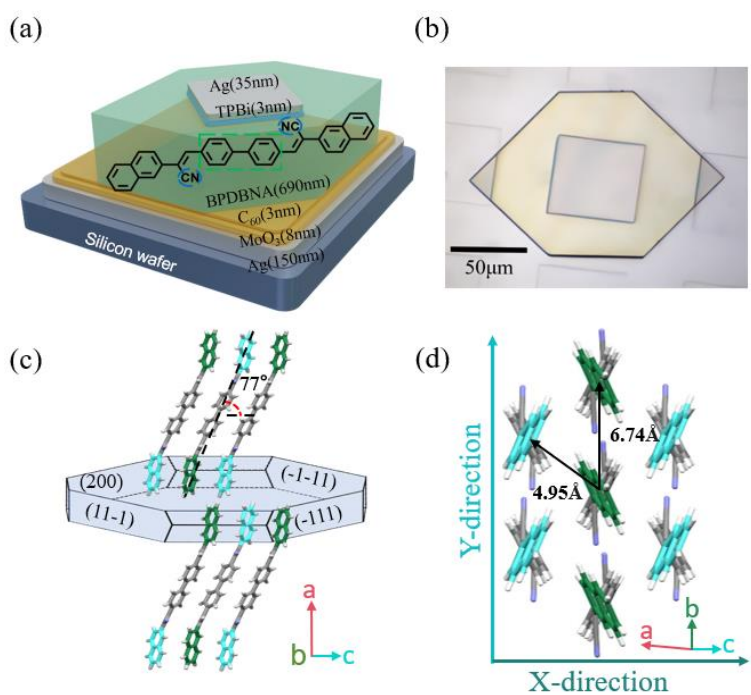

**Figure 2. OLED architecture, BPDBNA molecule and their stacking in the crystals.** (a) Schematic structure of the OLED based on BPDBNA single crystal and chemical structures of the BPDBNA molecule. (b) The optical image of the BPDBNA OLED. (c)

Single-crystal cell-unit structure of BPDBNA. (d) A scheme of BPDBNA crystals and molecular stacking arrangement.

The synthesis and characterization of BPDBNA molecule and its single microcrystals are detailed described in the Supplementary Materials (Figure S2-S7 and Table S1). Figure 2a shows a schematic of the single-crystal OLED, which were fabricated with a top-emitting device architecture: silicon wafer/Ag (150 nm)/MoO$_3$ (8 nm)/C$_{60}$ (3 nm)/BPDBNA microplate (690 nm)/TPBi (3 nm)/Ag (35 nm). From the bright-field optical image of this device (Figure 2b), two metallic silver layers are separated by the organic layer of BPDBNA microplate. The bottom and top Ag films played the role of hole and electron injection layers, respectively. Because of the excellent reflective properties of the silver films (that is, the reflectivity of the silver film with the thickness of 150 nm is more than 99% and that of 35 nm is about 50%), the parallel silver electrodes also operate as high-quality reflectors forming an optical Fabry-Pérot microcavity.

According to the experimental results of cyclic voltammetry (CV) test and ultraviolet photoelectron spectroscopy (UPS) test, the highest occupied molecular orbital (HOMO) and LUMO of BPDBNA were determined to be -6.10 eV and -3.50 eV, respectively (Figure S8). Notably, the deep LUMO of -3.50 eV brings about two advantages, i) high electron affinity reduces the non-radiative recombination losses due to exciton quenching and trap-assisted recombination[32]. ii) The non-reactive electron-injection layer can be used to increase the air stability of the OLEDs. Therefore, we chose TPBi as the electron transport layer to reduce the electron injection barrier. Especially, we used high-work-function transition metal oxides (MoO$_3$) in combination with C$_{60}$ as a barrier layer to solve hole injection challenge for the organic materials with deep HOMO of -6.10 eV[28,29,31].

The BPDBNA microplates exhibit strong emission with a photoluminescence (PL) quantum yield of PLQY = 78 ± 5% obtained through an absolute method by using an integration sphere. To understand the efficient luminescence mechanism, we analyzed the molecular arrangement in the crystals and theoretically calculated the relevant photophysical parameters (details in Supplemental Materials). The structural characterizations indicate that the BPDBNA molecule adopts herringbone packing within the *bc* layer (Figure S9-S10) and forms an H-type aggregation. In comparison with the single-crystal cell-unit structure, BPDBNA molecules in the microplate stack as a lamellar arrangement along the *a*-axis (Figure 2c-d), with the layer height $a$ = 54.1 Å corresponding to the molecule length 26.7 Å. The calculated excitonic couplings (J) accounting for the Coulomb dipole-dipole interactions at B3LYP/6-31G(d) level via NWchem6.3 packages[33] support the assertion of H-type coupling[34]. Notably, H-aggregation has demonstrated to endow the materials with strong emission, balanced charge transports and giant optical anisotropy[35]. The calculated electronic coupling V and reorganization energy $E_{reorg}$ (TableS4), acquired at B3LYP/6-31G(d) level by interfacing MOMAP4 transport module and the Gaussian09 Program[36-40], indicate that BPDBNA possesses a balanced ambipolar transport. Indeed, the calculated results show that BPDBNA exhibits balanced hole and electron mobilities of $\mu_h$ = 0.012 cm$^2$ V$^{-1}$ s$^{-1}$ and $\mu_e$ = 0.030 cm$^2$ V$^{-1}$ s$^{-1}$, respectively. We also measured the refractive indices (*n*) along the different directions and determined them as $n_x$ = 1.7 and $n_y$ = 3.1, respectively

(Figure S14-S16), which testifies the giant optical anisotropy in BPDBNA microcrystals.

**Spin-valley locking from photonic SOC**

As mentioned above, the herringbone molecular packing arrangement endows BPDBNA microcrystals with a (giant) optical anisotropy (birefringence).[26] This refractive index anisotropy can be described by the projection of the ordinary ($n_o$) and extraordinary ($n_e$) refractive indices in the XY plane (Figure 2d). At normal incidence, for a cavity mode with $l$ antinodes, two orthogonally linearly polarized modes, denoted by $X_l$ and $Y_l$, are therefore separated in energy due to the difference of the effective refractive indices.

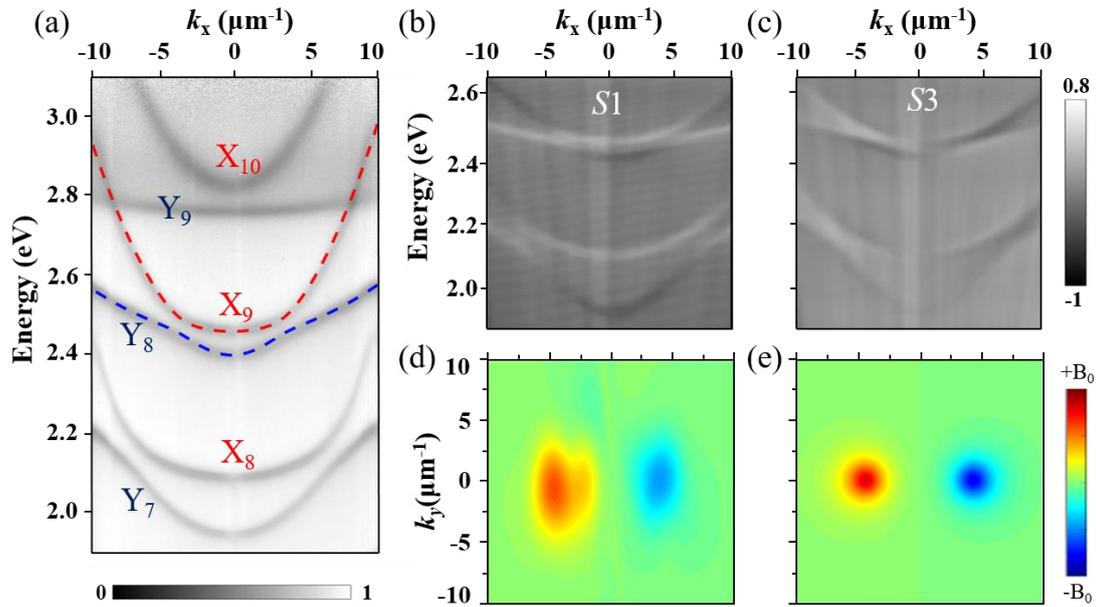

**Figure 3. Angle-resolved reflectivity spectrum and Stokes parameters.** (a) ARR of the selected sample by employing a halogen lamp as the unpolarized white-light source. Measured Stokes parameters $S1$ (b) and $S3$ (c) of $X_9$, $Y_8$, $X_8$ and $Y_7$ modes. Experimental Berry curvature $B_z$ extracting from the measured Stokes vector and theoretically simulated Berry curvature $B_z$ for the $X_9$ mode (d,e).

Figure 3a shows the angle-resolved reflectivity (ARR) of our selected device with a typical thickness of about 690 nm. Clearly, two sets of modes with distinctive curvatures (masses) can be seen, with crossings and anticrossings determined by their parity. We focus on a specific pair of modes of opposite parity exhibiting anticrossing: $X_9$ and $Y_8$. Fitting their dispersion with the effective Hamiltonian (1) allows us to determine all its parameters: the effective mass $m = (1.35 \pm 0.03) \times 10^{-5} m_0$ where $m_0$ is the free electron mass, the XY splitting $\beta_0 = 60 \pm 5$ meV and $\beta_2 = 0.94 \pm 0.03$ meV μm², the TE-TM splitting $\beta_1 = 2.4 \pm 0.1$ meV μm², and the emergent OA $\alpha = 10 \pm 2$ meV μm.

To determine the polarization characteristics of these modes, the polarization-resolved ARR experiments have been carried out for the six different polarization components of light (horizontal-vertical, diagonal-anti-diagonal, and left and right circular) to construct 3D tomography. To analyze the contribution of the birefringence, we have calculated these two modes by using the 2D cavity photon dispersion relations (details in Supplementary Materials)[25]. The simulated refractive indices of two modes are 1.7 for X-polarized modes and 3.1 for Y-polarized ones, respectively, which match well with the experimental results (Figure S16).

As shown in Figure 3a, the clear splitting between linearly polarized modes (such as $X_{10}$ and $X_9$) at $k_x = 0$ becomes comparable with that of the modes of different orders (such as $X_9$ and $Y_8$) because of the significant optical anisotropy of BPDBNA microcrystals measured above. The $X_9$ and $Y_8$ modes anticross at a finite wave vector (instead of simply crossing) and generate two gapped tilted Dirac cones at the two reciprocal space points. We experimentally measured the Stokes vector of these modes

(details in Supplementary Materials). The linear polarization of the $X_9$ branch changes signs across the anticrossing regions (Figure 3b). In comparison, the circularly polarized signals in the $X_9$ branch reach the maximal values at the anticrossing points and change sign at $k_x = 0$ because of the time-reversal symmetry (Figure 3c). Figure S17 show the 2D Stokes vector of the $X_9$ branch in the reciprocal space. The values of the $S1$ component are zero around two anticrossing points, while the value of the $S3$ component reaches the maximum in the same regions, but with opposite signs. The measurements of the Stokes vector allow extracting the Berry curvature of the modes. As shown in Figure 3d, the Berry curvature for $X_9$ branch shows two maxima of opposite signs separating the reciprocal space in two regions, topologically nontrivial valleys[23,26,41]. They are characterized by opposite valley Chern numbers equal to $\pm 1/2$. The experimentally extracted Berry curvature (Figure 3d) corresponds very well to the theoretical predictions (Figure 3e) based on the 2 × 2 effective Hamiltonian (1) with parameters extracted from the experimental dispersion, as indicated above (Figure 3a). We also measured the angle-resolved (AR) PL spectrum of this device (Figure S18). Notably, the spin-valley-locked CP PL emission is observed from the two anticrossing points, similar to that in the above reflectivity.

**EL Performances of the CP-OLEDs**

The EL measurements were performed using a homemade setup as shown in in Scheme S2. Figure 4a shows the energy level diagram of our microcavity CP-OLED architecture, where Ag/MoO$_3$/C$_{60}$ and Ag/TPBi are used as the lower and upper electrodes to inject holes and electrons, respectively. The EL measurement setup is shown in detail in Scheme S2. Prior to characterizing the EL performances of our devices, we designed and measured pure-hole devices and pure-electron devices based on BPDBNA microcrystals. According to the results of the J-V curves of the pure-carrier devices (Figure S19), it is clearly seen that the hole injection and electron injection are both effective in the device architectures.

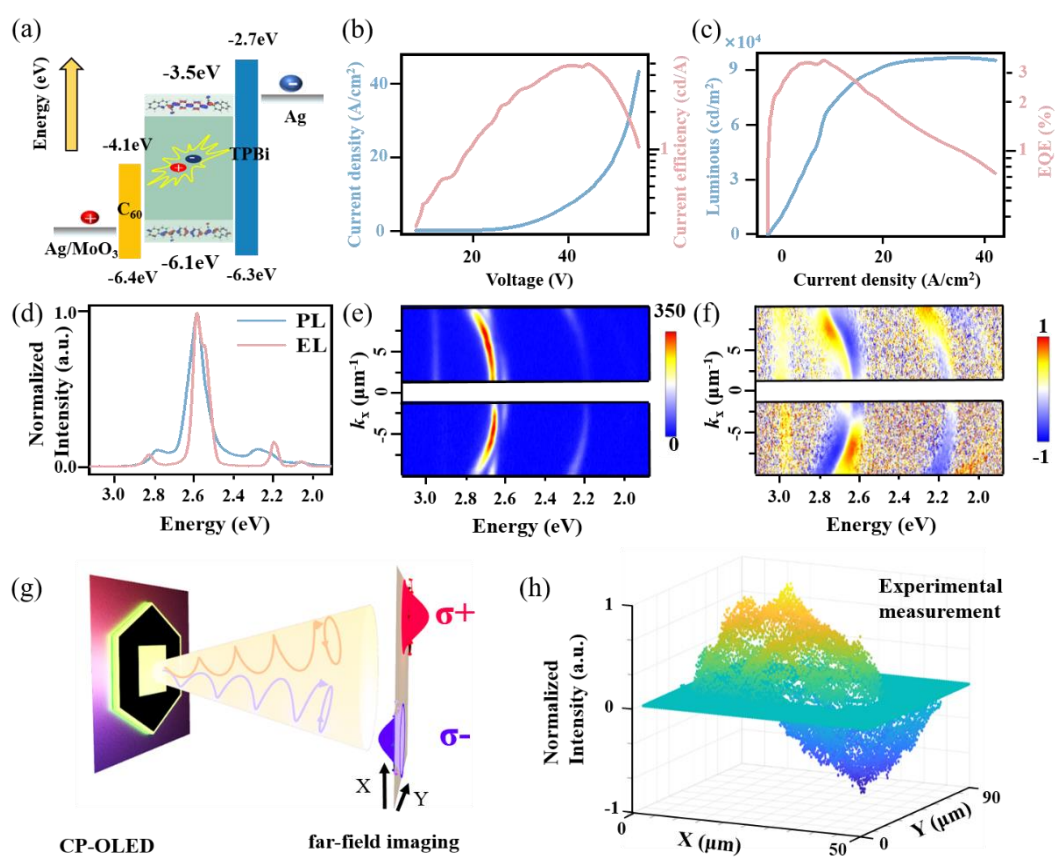

**Figure 4**. **Device characteristics and EL emission.** (a) Energy level diagram of the device. (b) Dependence of the current density (blue line) and the current efficiency (red line) on the voltage. (c) Dependence of the luminescence (blue line) and the EQE (red line) on the current density. (d) EL and PL spectra of the OLEDs. (e) Polarized angle-resolved EL spectra. (f) Stokes vector $S3$ of this device. Schematic (g) and experimental measurement (h) of far-field imaging of CP-OLED.

The EL performances of the BPDBNA CP-OLEDs are summarized based on the measured results of beyond 50 devices. Figure 4b shows the curves of the current density and current efficiency as a function of voltage. The current density of the devices is as high as ~ 45 A/cm$^2$ because of the good crystal quality of BPDBNA microcrystals, which suggests that our OLED can withstand high voltages up to tens of Volts and stably emit EL without any damage. The current efficiency is also calculated to be about 5.1 cd/A. Thanks to the near-barrier-free carrier injection and balanced transport, the highest luminance and EQE of our OLED reaches nearly 98000 cd/m$^2$ and 3.65% (Figure 4c). Notably, EQE = $\Phi_s \times \chi \times \eta_{out}$ was widely adopted for OLEDs, where $\Phi_s$ is BPDBNA PLQY of 0.78, $\chi$ is the fraction of spin statistics of 25% for fluorescence materials in the EL process[42], and $\eta_{out}$ is light outcoupling factor of usually 20%. The theoretical upper limit EQE for this BPDBNA device can be estimated as 3.9%. EQE of 3.65% for our CP-OLEDs is already approaching its limit value.

Under a certain voltage, the device achieves uniform and bright EL emission from the surface and waveguided emission from the edges (Figure S20). It indicates that the EL emission comes from the active layer BPDBNA of the OLED and is regulated by the microcavity. Figure 4d shows the EL spectrum of this device. The EL spectra present a narrowband emission with full width at half maximum (FWHM) of 16 nm. To study the CP property of the EL emission, we performed the angle-resolved EL (AREL) measurement for our OLED. The angle range of the AREL spectrum is only ±15°, limited by numerical aperture of the microscope lens in our setup. Expectedly, the obtained AREL spectra also show clear left-handed and right-handed (σ+ and σ−) CP emissions from the two anticrossing points as shown in Figure 4e and 4f, which is agreement with the results of the ARPL (Figure S21). This strongly testifies that the photonic SOC is still present under electrical excitation and that it induces the spin-valley-locking leading to CP emission. In fact, this spin-valley-locked CP EL can be observed from multiple wavelengths in our devices. The key parameters of CP asymmetry factor ($g_{EL}$ for EL and $g_{lum}$ for PL) are usually employed to characterize the performance of CP-OLEDs and can be defined as $g_{EL}$ (or $g_{lum}$) = $2\times(I_L - I_R)/(I_L + I_R)$, where $I_L$ and $I_R$ correspond to the intensities of left- and right-handed polarization, respectively[43]. In our structure, $g_{lum}$ and $g_{EL}$ reach the maximum values of about 1.76 (Figure S18) and 1.80, respectively.

In order to further commercialize the device, near-field imaging of the spin-valley-locked emission was carried out for the CP-OLEDs, as illustrated in Figure 4g. Experimentally, the clear spin-valley-locked circular polarized-emissions at the different positions in the real space have been observed (Figure S22). Figure 4h further demonstrates the separation (~20 μm) of left- and right-handed circularly polarized light in real space. Hence, our spin-valley-locked CP-OLEDs have exhibited the best EL performances, including luminance, EQE and $g_{EL,}$ among the currently available single-crystal CP-OLEDs.

## Conclusion

In summary, we propose a distinct strategy for designing spin-valley-locked single-crystal CP-OLEDs with high luminance, EQE and large $g_{EL}$ through introducing artificial photonic SOC. Based on the design of a novel organic molecule, BPDBNA, with balanced charge injection and transport properties, our chiral-emitter-free CP-OLEDs exhibit an narrowband emission of 16 nm, EQE of 3.65%, a maximum luminance of near 98000 cd/m$^2$ and a g$_{EL}$ of up to 1.80, which are among the best performances of active single-crystal CP-OLEDs. The EL with strong opposite circular polarizations is emitted selectively from two optical valleys and their emission angles are respectively 20º and -20º. These unique spin-valley-locked CP-OLEDs with top-emitting architecture demonstrate a promising strategy for light sources, display technologies and spin photonic applications.

# Reference

1   Carr, R., Evans, N. H. & Parker, D. Lanthanide complexes as chiral probes exploiting circularly polarized luminescence. *Chem. Soc. Rev.* **41**, 7673-7686 (2012).
2   Sang, Y., Han, J., Zhao, T., Duan, P. & Liu, M. Circularly polarized luminescence in nanoassemblies: generation, amplification, and application. *Adv. Mater.* **32**, 1900110 (2020).
3   Zhang, Y.-P. *et al.* Chiral spiro-axis induced blue thermally activated delayed fluorescence material for efficient circularly polarized oleds with low efficiency roll-off. *Angew. Chem. Int. Ed.* **60**, 8435-8440 (2021).
4   Yoshida, K., Gong, J., Kanibolotsky, A. L., Turnbull, G. A. & Samuel, I. D. W. Electrically driven organic laser using integrated OLED pumping. *Nature* **621**, 746-752 (2023).
5   Wan, L. *et al.* Inverting the handedness of circularly polarized luminescence from light-emitting polymers using film thickness. *Acs Nano* **13**, 8099-8105 (2019).
6   Wan, L., Liu, Y., Fuchter, M. J. & Yan, B. Anomalous circularly polarized light emission in organic light-emitting diodes caused by orbital–momentum locking. *Nat. Photon.* **17**, 193-199 (2022).
7   Furlan, F. *et al.* Chiral materials and mechanisms for circularly polarized light-emitting diodes. *Nat. Photon.* DOI: 10.1038/s41566-024-01408-z, (2024).
8   Ma, X.-H. *et al.* Carbene-stabilized enantiopure heterometallic clusters featuring EQE of 20.8% in circularly-polarized OLED. *Nat. Commun.* **14**, 4121 (2023).
9   Feuillastre, S. *et al.* Design and synthesis of new circularly polarized thermally activated delayed fluorescence emitters. *J. Am. Chem. Soc.* **138**, 3990-3993 (2016).
10  Li, M., Wang, Y.-F., Zhang, D., Duan, L. & Chen, C.-F. Axially chiral TADF-active enantiomers designed for efficient blue circularly polarized electroluminescence. *Angew. Chem. Int. Ed.* **59**, 3500-3504 (2020).
11  Gong, Z.-L. *et al.* Frontiers in circularly polarized luminescence: molecular design, self-assembly, nanomaterials, and applications. *Sci. China Chem.* **64**, 2060-2104 (2021).

12   Liu, Z.-F. *et al.* Circularly polarized laser emission from homochiral superstructures based on achiral molecules with conformal flexibility. *Angew. Chem. Int. Ed.* **62**, e202214211 (2023).

13   Song, F. *et al.* Highly efficient circularly polarized electroluminescence from aggregation-induced emission luminogens with amplified chirality and delayed fluorescence. *Adv. Funct. Mater.* **28**, 1800051 (2018).

14   Geng, Z., Zhang, Y., Zhang, Y., Quan, Y. & Cheng, Y. Amplified circularly polarized electroluminescence behavior triggered by helical nanofibers from chiral co-assembly polymers. *Angew. Chem. Int. Ed.* **61**, e202202718 (2022).

15   An, M.-H. *et al.* Highly polarized emission from organic single-crystal light-emitting devices with a polarization ratio of 176. *Optica* **9** (2022).

16   Xiong, J., Hsiang, E.-L., He, Z., Zhan, T. & Wu, S.-T. Augmented reality and virtual reality displays: emerging technologies and future perspectives. *Light Sci. Appl.* **10**, 216 (2021).

17   Wang, Q. *et al.* Constructing highly efficient circularly polarized multiple-resonance thermally activated delayed fluorescence materials with intrinsically helical chirality. *Adv. Mater.* **35**, 2305125 (2023).

18   Deng, Y. *et al.* Circularly polarized luminescence from organic micro-/nano-structures. *Light Sci. Appl.* **10**, 76 (2021).

19   Terças, H., Flayac, H., Solnyshkov, D. D. & Malpuech, G. Non-abelian gauge fields in photonic cavities and photonic superfluids. *Phys. Rev. Lett.* **112**, 066402 (2014).

20   Gianfrate, A. *et al.* Measurement of the quantum geometric tensor and of the anomalous Hall drift. *Nature* **578**, 381-385 (2020).

21   Lu, L., Joannopoulos, J. D. & Soljačić, M. Topological photonics. *Nat. Photon.* **8**, 821-829 (2014).

22   Ozawa, T. *et al.* Topological photonics. *Rev. Mod. Phys.* **91**, 015006 (2019).

23   Ren, J., Liao, Q., Li, F., Fu, H. & Solnyshko, D. Nontrivial band geometry in an optically active system. *Nat. Commun.* **12**, 689-969 (2021).

24   Whittaker, C. E. *et al.* Optical analogue of dresselhaus spin–orbit interaction in photonic graphene. *Nat. Photon.* **15**, 193-196 (2020).

25   Ren, J. *et al.* Realization of exciton-mediated optical spin-orbit interaction in organic microcrystalline resonators. *Laser Photon. Rev.* **16**, 2100252 (2022).

26   Long, T. *et al.* Helical polariton lasing from topological valleys in an organic crystalline microcavity. *Adv. Sci.* **9**, 2203588 (2022).

27   Jia, J. *et al.* Circularly polarized electroluminescence from a single-crystal organic microcavity light-emitting diode based on photonic spin-orbit interactions. *Nat. Commun.* **14**, 31 (2023).

28   Kotadiya, N. B., Blom, P. W. M. & Wetzelaer, G.-J. A. H. Efficient and stable single-layer organic light-emitting diodes based on thermally activated delayed fluorescence. *Nat. Photon.* **13**, 765-771 (2019).

29   Kotadiya, N. B., Wetzelaer, G.-J. A. H., Lu, H., Mondal, A. & Ie, Y. Universal

strategy for ohmic hole injection into organic semiconductors with high ionization energies. *Nat. Mater.* **17**, 329-334 (2018).
30  Rechcińska, K. *et al.* Engineering spin-orbit synthetic hamiltonians in liquid-crystal optical cavities. *Science* **366**, 727-730 (2019).

31  Nicolai, H. T. *et al.* Unification of trap-limited electron transport in semiconducting polymers. *Nat. Mater.* **11**, 882-887 (2012).

32  Kuik, M., Koster, L. J. A., Dijkstra, A. G., Wetzelaer, G. A. H. & Blom, P. W. M. Non-radiative recombination losses in polymer light-emitting diodes. *Org. Electron.* **13**, 969-974 (2012).

33  Jiang, Y., Shuai, Z. & Liu, M. Roles of long-range hopping, quantum nuclear effect, and exciton delocalization in exciton transport in organic semiconductors: a multiscale study. *J. Phys. Chem. C* **122**, 18365-18375 (2018).

34  Spano, F. C. Excitons in conjugated oligomer aggregates, films, and crystals. *Ann. Rev. Phys. Chem.* **57**, 217-243 (2006).

35  KASHA, M., RAwLs, H. R. & EL-BAYOUMI, M. A. THE exction model in molecular spectroscopy. *Pure Appl. Chem.* **11**, 371-392 (1965).

36  Marcus, R. A. Electron transfer reactions in chemistry. Theory and experiment. *Rev. Mod. Phys.* **65**, 599-610 (1993).

37  Niu, Y. *et al.* MOlecular materials property prediction pckage (MOMAP) 1.0: a software package for predicting the luminescent properties and mobility of organic functional materials. *Mol. Phys.* **116**, 1078-1090 (2018).

38  Nan, G. J., Yang, X. D., Wang, L. J., Shuai, Z. G. & Zhao, Y. Nuclear tunneling effects of charge transport in rubrene, tetracene, and pentacene. *Phys. Rev. B* **79**, 115203 (2009).

39  Shuai, Z., Wang, D., Peng, Q. & Geng, H. Computational evaluation of optoelectronic properties for organic/carbon materials. *Acc. Chem. Res.* **47**, 3301-3309 (2014).

40  Shuai, Z., Geng, H., Xu, W., Liao, Y. & Andre, J.-M. From charge transport parameters to charge mobility in organic semiconductors through multiscale simulation. *Chem. Soc. Rev.* **43**, 2662-2679 (2014).

41  Gao, F. *et al.* Topologically protected refraction of robust kink states in valley photonic crystals. *Nat. Phys.* **14**, 140-144 (2017).

42  Baldo, M. A., Holmes, R. J. & Forrest, S. R. Prospects for electrically pumped organic lasers. *Phys. Rev. B* **66**, 035321 (2002).

43  Sun, C.-L. *et al.* Lasing from an organic micro-Helix. *Angew. Chem. Int. Ed.* **59**, 11080-11086 (2020).

**Acknowledgements**

This work was supported by the National Natural Science Foundation of China (Grant Nos. 22150005, 22090022, and 22275125), the National Key R&D Program of China (2022YFA1204402, 2018YFA0704805, and 2018YFA0704802), the Natural Science

Foundation of Beijing, China (KZ202110028043), R&D Program of Beijing Municipal Education Commission (KM202210028016, BPHR202203119), the Science and Technology Innovation Program of Hunan Province (2022RC4039), Beijing Advanced Innovation Center for Imaging Theory and Technology. Additional support was provided by the ANR Labex GaNext (ANR-11-LABX-0014), the ANR program "Investissements d'Avenir" through the IDEX-ISITE initiative 16-IDEX-0001 (CAP 20-25), the ANR projects "MoirePlusPlus" and "NEWAVE" (ANR-21-CE24-0019) and the European Union's Horizon 2020 program, through a FET Open research and innovation action under the grant agreements No. 964770 (TopoLight).

The authors thank Dr. H.W. Yin from ideaoptics Inc. for the support on the angle-resolved spectroscopy measurements. We gratefully acknowledge HZWTECH for providing computation facilities.

**Author contributions**

Y.B.D., T.L., P.Y.W., H.H., Z.J.D. and Q.L designed the experiments and performed experimental measurements. G.M. and D.S. performed theoretical calculations. C.L.G., C.B.A., B.L., H.B.F. and Q.L. wrote the manuscript with contributions from all authors. H.B.F. and Q.L. supervised the project. All authors analyzed the data and discussed the results.

**Competing interests**

The authors declare no competing interests.

**Additional information**

Correspondence should be addressed to C.L.G., D.S., H.B.F. and Q.L.:

clgu@ipe.ac.cn, dmitry.solnyshkov@uca.fr hbfu@cnu.edu.cn, liaoqing@cnu.edu.cn

# Supporting Information

**Spin-valley-locked Electroluminescence for High-performance Circularly Polarized Organic Light-Emitting Diodes**

Yibo Deng,[1,†] Teng Long,[1,†] Pingyang Wang,[1] Han Huang,[1] Zijian Deng,[1] Chunling Gu,[2,*] Cunbin An,[1] Bo Liao,[3] Guillaume Malpuech,[4] Dmitry Solnyshkov,[4,5*] Hongbing Fu,[1,3,*] Qing Liao[1,*]

[1]Beijing Key Laboratory for Optical Materials and Photonic Devices, Department of Chemistry, Capital Normal University, Beijing 100048, China

[2]Institute of Process Engineering, Chinese Academy of Sciences, Beijing, 100190, China

[3]School of Materials Science and Engineering, Hunan University of Science and Technology, Xiangtan, Hunan 411201, China

[4]Université Clermont Auvergne, Clermont Auvergne INP, CNRS, Institut Pascal, F-63000 Clermont-Ferrand, France

[5]Institut Universitaire de France (IUF), 75231 Paris, France

†These authors contributed equally: Yibo Deng, Teng Long.

## MATERIALS AND METHODS

### 1. Design of BPDBNA molecule

The energy level from theoretical calculations of BPDBNA molecule and the control molecule 4,4'-bis((*E*)-2-(napthalen-2-yl)vinyl)-1,1'-biphenyl (BPNVB)[1,2].

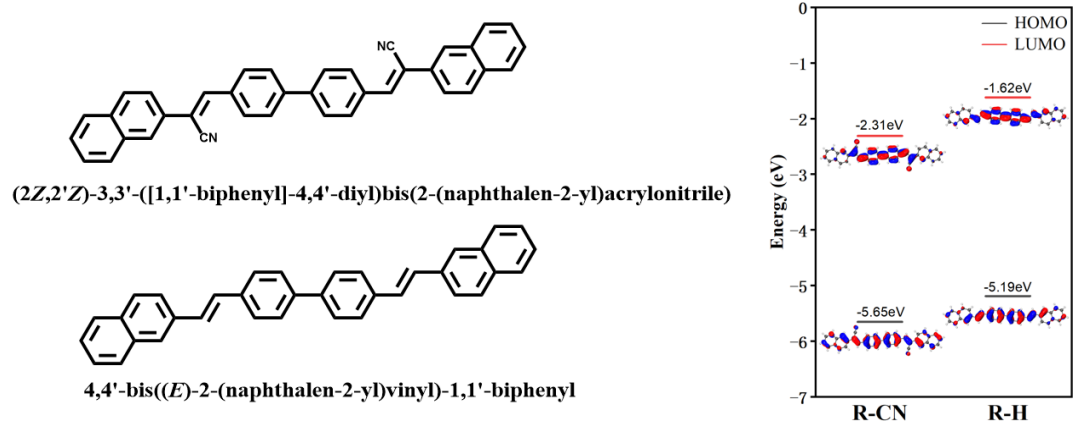

**Figure S1.** Theoretical calculations for the energy levels of BPDBNA and BPNVB molecules. It is clear that the LUMO energy of BPDBNA decease in comparison to the BPNVB due to the introduction of -CN group.

### 2. Synthesis of BPDBNA molecule

BPDBNA molecules were synthesis via a Horner-Wadsworth-Emmons reaction[3]. The skeleton molecules, 4,4'-Biphenyldicarboxaldehyde (420.46 mg, 2 mmol) and 2-Naphthylacetonitrile (668 mg, 4 mmol), were added to the branch tube reactor. Under the protection of argon (Ar), 20 mL ethanol and 1 mL ultradry tetrahydrofuran were injected with a syringe. After full reaction in the ice bath for 1 hour, the solution gradually changed from milky white light green. Then 112 mg of potassium tert-butanol was weighed and dissolved in ethanol and added to the branch tube reactor. The branch reactor was transferred to the oil bath, set the oil bath temperature to 65 °C and then close the oil bath pan after 4 hours. After cooling to room temperature, the solution was filtered and rinsed with H₂O, ethanol (C₂H₅OH) and dichloromethane (CH₂Cl₂) for 2-3 times, respectively. After drying on the hot table, the product was filled in the bottle, wrapped the bottle with tin foil and tie several holes. Finally, this bottle was put in a drying oven at 70 °C for 6 hours. The yellow powder was obtained as the target product, and the yield was 80%.

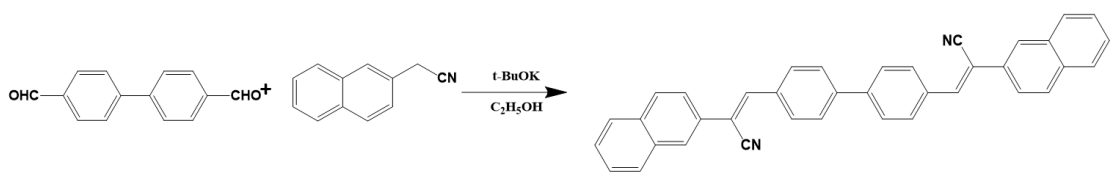

**Figure S2.** The synthesis of the BPDBNA molecule.

Because of poor solubility of BPDBNA, its $^1$H NMR and $^{13}$C NMR signals were not detected. The target compound was characterized using elemental analysis and high-resolution mass spectroscopy (MS) after the sublimation purification. HR-MS (MALDI-MS) m/z: 507.3 [M+]; Elemental analysis: Anal. calculated for $C_{38}H_{24}N_2$ (%): C: 89.14, H: 4.73, S: 5.51. Experimental: C: 84.76, H: 4.72, N: 5.51.

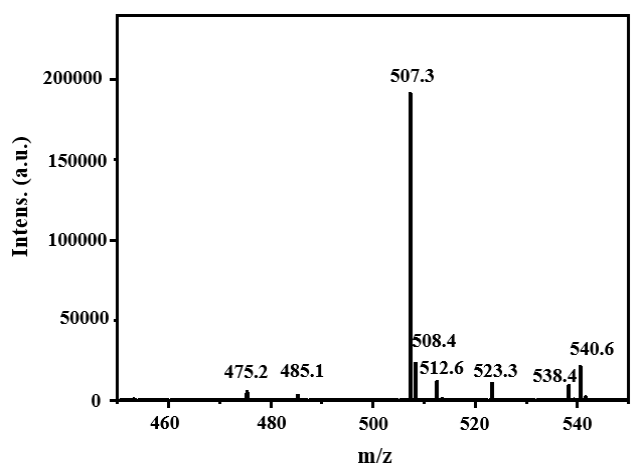

**Figure S3.** MALDI-MS spectrum of BPDBNA.

## 3. The preparation of BPDBNA single crystals

In our experiment, BPDBNA single crystals were fabricated using a facile physical vapor deposition (PVD) method. Typically, a quartz boat carrying 10 mg BPDBNA was placed in the center of a quartz tube which was inserted into a horizontal tube furnace. A continuous flow of cooling water inside the cover caps was used to achieve a temperature gradient over the entire length of the tube. To prevent oxidation of BPDBNA, Ar was used as inert gas during the PVD process (flowrate: 45 sccm·min$^{-1}$). The pre-prepared hydrophobic substrates were placed on the downstream side of the argon flow for product collection and the furnace was heated to the sublimation temperature of BPDBNA (at temperature region of ~ 280 °C), upon which it was physically deposited onto the pre-prepared hydrophobic substrates for 10 hours.

## 4. Methods

High-resolution mass spectroscopy was recorded at the analytical center of the Institute of Chemistry, Chinese Academy of Science (ICCAS). Elemental analysis was performed on Flash Smart, CHN-S elemental analyses. Thermal gravimetric analysis (TGA) was performed on a STA 409 PC thermo-gravimeter in air at a heating rate of 5 °C min$^{-1}$.

The crystals samples were characterized by transmission electron microscopy (TEM, JEM-1011, JEOL) were obtained by transferring two samples on a carbon-coated copper grid. TEM measurement was performed at room temperature at an accelerating voltage of 100 kV. X-ray diffraction (XRD) patterns were measured by a D/max 2400 X-ray diffractometer with Cu Kα radiation (λ = 1.54050 Å) operated in the 2θ range from 5 to 35°, by using the samples on a cleaned quartz substrate. The height of microcrystals was measured by atomic force microscopy (AFM, Bruker Multi-Mode 8-HR).

Crystallographic data (excluding structure factors) for the structure of BPDBNA reported in this paper were deposited in the Cambridge Crystallographic Data Centre (CCDC:2324127). The fluorescence and absorption spectra were obtained by a fluorescence spectrometer (Hitachi F-4600) and an ultraviolet-visible spectrophotometer (Hitachi U-3900H), respectively. The absolute fluorescence quantum yields were measured by using the Edinburgh FLS1000 with an integrating sphere. Bright-field optical images and fluorescence microscopy images were taken by a home-made optical microscopy equipped with a 50 × 0.9 NA excitation objective and a charge couple device (CCD, Olympus DP-73, Tokyo, Japan) at room temperature in air. The excitation source is a Xenon lamp equipped with a band-pass filter (330~380 nm for UV-light) and the samples were deposited onto a quartz slide.

Electrochemical measurements (CV) were performed with a CHI660C electrochemistry station in acetonitrile solution with sample concentration of $10^{-3}$ M using Bu4NPF6 as electrolyte, Pt as working electrode, platinum wire as auxiliary electrode, and a porous glass wick Ag/AgCl as pseudo-reference electrode. Ultraviolet photoelectron spectroscopic (UPS) was tested by vaporizing BPDBNA molecules on a 1 cm×1 cm ITO conductive substrate with a thickness of 15 nm through an organic vaporizer.

**5. Theoretical calculation**

The ground state and excited states were optimized by the density functional theory

(DFT) and time-dependent density functional theory (TD-DFT) at the B3LYP/6-31G(d) level in Guassian16 Program. A quantum mechanics/molecular mechanics (QM/MM) approach was employed to simulate the crystal stacking environment. This entailed the embedding of the monomers in the lattice surrounding the molecule, with the latter serving as a rigid MM part.

As a conventional DFT generalization, B3LYP is not able to represent the long-range weak interactions between molecules, so we used the DFT-D3 level of dispersion correction and calculated its dispersion energy. For mobility calculation, we have applied B3LYP functional with 6-31G(d) basis set for reorganization energy and B3LYP functional with 6-31G(d) basis set for transfer integral by Guassian09 Program. The charge transfer is calculated based on multimode quantum charge transfer rate by MOMAP.[4-8] Theoretical calculations and experimentally relevant parts of the data are shown in Figure S9-11 and Table S2-3.

## 6. Preparation and characterization of BPDBNA CP-OLEDs

The microcavity CP-OLEDs were fabricated with a top-emitting device architecture. First, a hole transport layer of 3 nm $C_{60}$ and 6 nm $MoO_3$ and 150 nm Ag were thermally evaporated on a substrate grown by the PVD method, it is then flipped over by UV curing adhesive. Subsequently, a copper mesh mask plate was placed over the crystal and it was thermally vaporized with an upper layer of 3 nm TPBi and 35 nm Ag.

## 7. Preparation of hole-only and electron-only devices:

Highly n-doped (100) Si wafers (0.05-0.2 U·cm) were successively cleaned with piranha solution (70/30 vol./vol. $H_2SO_4/H_2O_2$), deionized water and isopropanol, respectively, and then were dried by $N_2$ and $O_2$ plasma for 10 min. First, 150 nm Ag, 6 nm $MoO_3$, and 3 nm $C_{60}$ were sequentially thermally deposited on the wafer, then transferred the crystal onto it and covered the copper mesh mask on top of the crystal. Then 3 nm $C_{60}$ and 6 nm $MoO_3$ and 100 nm Ag were thermally vaporized as pure hole devices. A 10 nm TPBi was thermally evaporated on N-Si, then the crystal was transferred on top and a copper mesh mask was placed on top of the crystal. Then 3 nm TPBi and 35 nm Ag were thermally vaporized as pure electronics.

**8. Spectroscopic and the angle-resolved spectroscopy characterization**

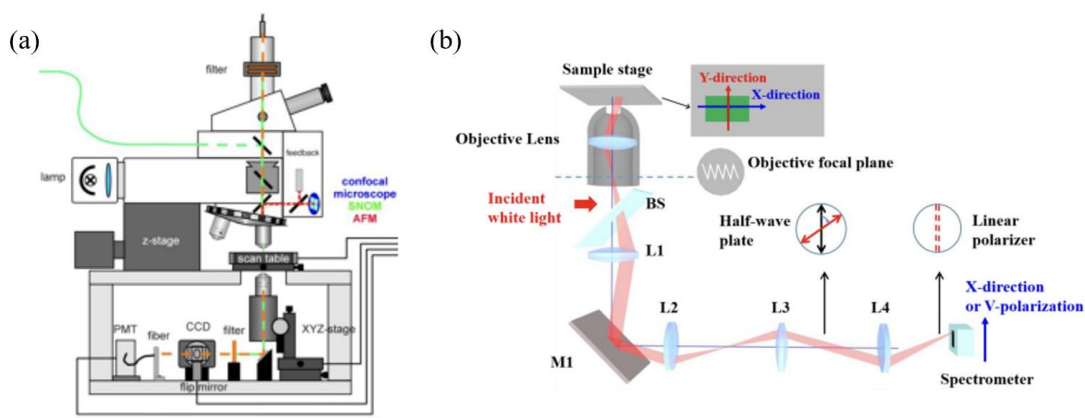

**Scheme S1.** (a) Schematic demonstration of the experimental setup for the optical characterization: the near-field scanning optical microscopy. (b) Experimental setup for the polarization-resolved PL. BS: beam splitter; L1-L4: lenses; M1: mirror. The red beam traces the optical path of the reflected light from the sample at a given angle.

The photoluminescence spectrum of the device was characterized by using a homemade optical microscope equipped with a 50 × 0.9 NA objective (Scheme S1a). A single selected device is excited on a two-dimensional (2D) movable table using a continuous laser focused at 405 nm to a 50-μm diameter spot. Spatially resolved PL spectra were collected underneath by using a 3D-movable objective and detected using a liquid-nitrogen cooled charge-coupled device (CCD). The angle-resolved spectroscopy was performed at room temperature by the Fourier imaging using a 100× objective lens of a NA 0.95, corresponding to a range of collection angle of ±60° (Scheme S1b).

The electroluminescence spectrum of the device (Scheme S2) was characterized by using a homemade optical microscope equipped with a 50 × 0.46 NA objective, corresponding to a range of collection angle of ±15°. Voltage was applied to the device through a Keithley 2400 in a glove box filled with a nitrogen atmosphere. Spatially resolved spot spectra were collected on top using a three-dimensional movable objective and detected using an electrically cooled charge-coupled device (CCD).

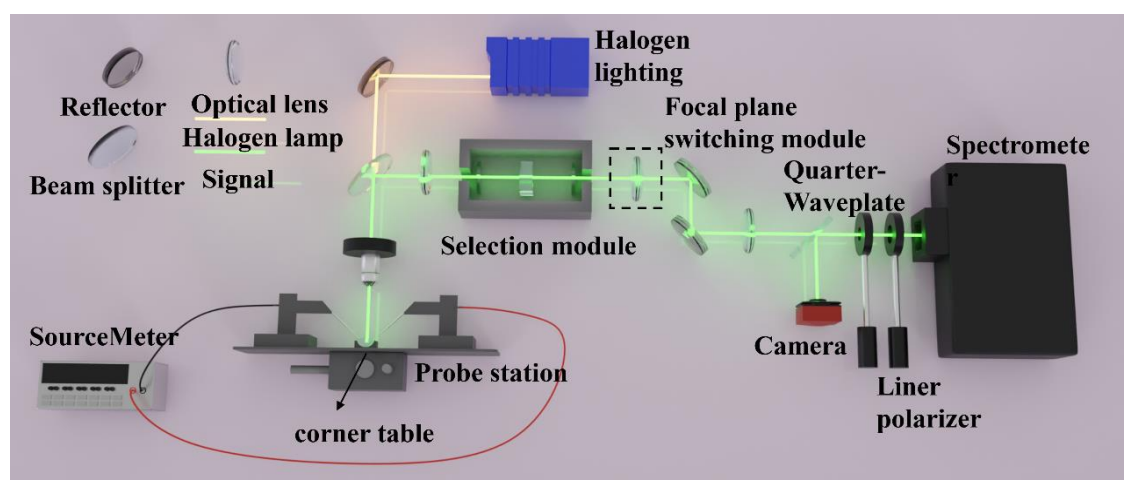

**Scheme S2**. Scheme of tailor-made device for recording emission image, spectra and current simultaneously. For maintain stability, the whole set of equipment is placed on the optical platform. A source-meter (Keithley 2400) equipped with probes is applied to electrical measurement. The yellow line represents calibration light path origin from a halogen lamp with a known radiation flux $Pc(\lambda)$. The green line is collection and observation path.

The incident white light of the Halogen lamp with the wavelength of 400-700 nm is used to focus on the region to be measured. The k-space or angular distribution of the reflected light was located at the back focal plane of the objective lens. In addition, the tilt angle of the device can be changed to acquire angle-resolved spectra at larger angles by means of an angular stage.

In order to investigate the polarization properties, we placed a linear polarizer, a half-wave plate and a quarter-wave plate in front of spectrometer to obtain the polarization state of each pixel of the *k*-space images in the horizontal-vertical (0° and 90°) and circular (σ+ and σ−) basis.

## 9. Simulation of cavity modes in reflection spectrum and refractive index of BPDBNA single crystal

We have calculated X- and Y-polarized modes by using the two-dimensional cavity photon dispersion relations. According to the equation,

$$E_{\text{CMn}}(\theta) = \sqrt{\left(E_c^2 * \left(1 - \frac{\sin^2\theta}{n_{eff}^2}\right)^{-1}\right)} - (n-1)*l \qquad (5)$$

where $\theta$ represents the incidence angle, $E_{\text{CMn}}(\theta)$ is the cavity photon energy of the nth cavity mode as a function of $\theta$, Ec represents the cavity modes energy at $\theta = 0°$, $E_{\text{CM1}}(\theta)$ represents the energy of the first cavity mode when $n = 1$, $(n - 1) \times l$ represents the energy difference from the first cavity mode. The refractive indices are calculated to be 1.7 and 3.1 the red and black curves, respectively.

We measured the refractive index using the method in the reference[9]. Firstly, the BBPDBNA microbelts are placed on the quartz substrate, and the Fabry-Pérot interference is formed on the top and bottom smooth surfaces. The reflection spectra of the samples are measured with polarized light on parallel to and perpendicular to the long direction of the crystal. The reflection spectra of the samples are measured with polarized light on parallel to and perpendicular to the long direction of the crystal. The interference conditions are given by $2n(\lambda)d = m\lambda$, where $n(\lambda)$ is the refractive index at wavelength $\lambda$, m is the order of interference, and d is the crystal thickness. In the reflection spectra, the interference minimum occurs when m is an integer and the maximum occurs when m is a half integer. The wavelengths of the interference maximum and minimum were extracted from the reflection spectra measured when the polarization of incident light parallel to and vertical to the long direction of microbelts and are plotted as a function of crystal thickness. The black (red) lines are fitted by the black (red) points. Since two adjacent black (or red) lines have difference 1 in the interference order, the order m can be determined by $m = d_1/(d_2-d_1)$, where $d_1$ and $d_2$ ($d_2 > d_1$) are the thicknesses corresponding to the order m and $m+1$ respectively, at a fixed wavelength $\lambda$.

## 9. Calculation of EQE[10]

The electrically-driven light emission from OLEDs was detected by tailor-made equipment for recording emission photo, spectra and current simultaneously.

Formulaically speaking, according to the following definition equation (1), the EQE of EL device is acquired from the number of the collection emissive photons $n_v$ divided by the number of injected carriers $n_e$:

$$\text{EQE} = \frac{n_v}{n_e} \quad (1)$$

A halogen lamp with a known irradiation power of $P_c(\lambda)$, which calibrated by the Shanghai Institute of Measurement and Testing Technology, is used as the calibration light source. A reflector with a known reflectivity of R($\lambda$) is placed at the focal plane of the lens, and the light intensity by a CCD camera (PIXIS 100BR) equipped spectrometer after being reflected by the reflector is $I_c(\lambda)$, the response function of the light path to light of different wavelengths is $H(\lambda) = I_c(\lambda)/P_c(\lambda) \times R(\lambda)$. When testing electroluminescence samples, the obtained irradiation power at a specific wavelength is $P_s(\lambda) = I_s(\lambda)/H(\lambda)$, where $I_s(\lambda)$ is the electroluminescence intensity measured by the same detection light path and the spectrometer.

The distribution of photon number $n_v$, is calculated by equation (2). Hence, the number of total photons $n_v$ from detected emissive spectra ranging from starting and ending wavelength can be calculated through mathematical integral.

$$n_v = \int \frac{P_s(\lambda)\lambda\, d\lambda}{hc} \qquad (2)$$

where $n_v(\lambda)$ is the number of photons at any wavelength, in counts. $\lambda$ is wavelength of the emissive light, in nm. $h$ is Planck constant (6.626 × 10$^{-34}$ J s). $c$ is velocity of light in vacuum (3.0 × 10$^8$ m s$^{-1}$). The number of injected carriers ne during the test is calculated from the recombination current $I$ at saturation region of OLED devices divided by the elementary charge $e$ = 1.602 × 10$^{-19}$ C and integral time.

Therefore, the EQE is calculated by: $\eta_{EQE} = \dfrac{\int P_s(\lambda) * \frac{\lambda}{hc} d\lambda}{A/e} \qquad (3)$

Remarkably, our measurement system could only get only one surface emission or edge emission for a regular crystal during once test period. So the emission signal from the face in contact with the electrode and the crystal edge waveguide emission signal cannot be detected at the same time. The total EQE of the output light for an electroluminescence device was the summary of three EQEs from one surface emission and edge emissions.

$$EQE_{total} = EQE_{surface} + EQE_{edge} \qquad (4)$$

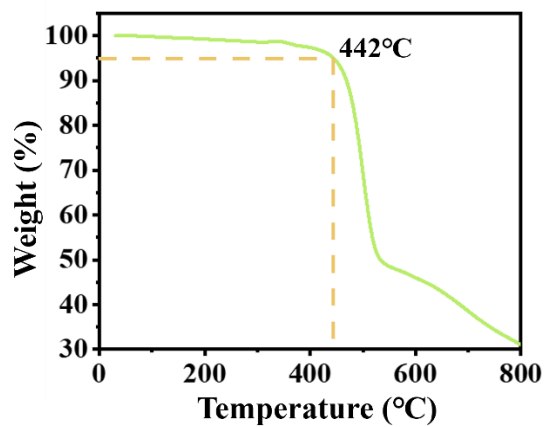

**Figure S4.** Thermal gravimetric analysis (TGA) of BPDBNA, showing that the thermal decomposition temperature is 442 °C. This decomposition temperature indicates the good thermal stability of BPDBNA.

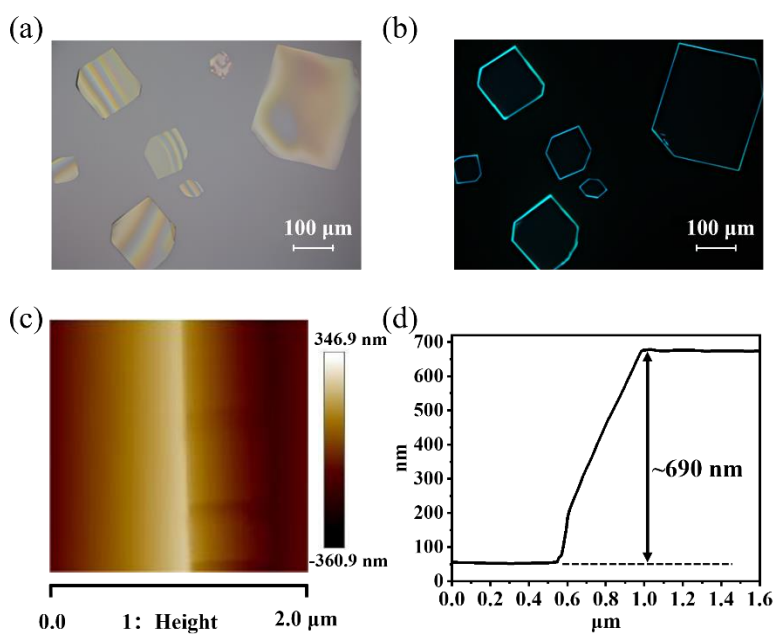

**Figure S5.** (a) and (b) Optical images of the different sizes of BPDBNA crystals. These BPDBNA microcrystals present a hexagonal plate-like structure with smooth outer surfaces and sharp edges. (c) and (d) Atomic force microscopy (AFM) image of BPDBNA crystals. The typical lateral size can be tens of micrometers and the thickness determined to be in the range of 400-1000 nm.

**Table S1. Crystal data and structure refinement for BPDBNA.**

| Identification code | BPDBNA |
|---|---|
| Empirical formula | $C_{38}H_{24}N_2$ |
| Formula weight | 508.59 |
| Temperature/K | 153.15 |
| Crystal system | monoclinic |
| Space group | C2/c |
| a/Å | 54.1487(13) |
| b/Å | 6.71400(10) |
| c/Å | 7.2763(2) |
| α/° | 90 |
| β/° | 92.496(2) |
| γ/° | 90 |
| Volume/Å$^3$ | 2642.82(10) |
| Z | 4 |
| $\rho_{calc}$g/cm$^3$ | 1.278 |
| μ/mm$^{-1}$ | 0.572 |
| F(000) | 1064.0 |
| Crystal size/mm$^3$ | 0.15 × 0.11 × 0.09 |
| Radiation | Cu Kα (λ = 1.54184) |
| 2Θ range for data collection/° | 9.81 to 148.39 |
| Index ranges | -59 ≤ h ≤ 67, -8 ≤ k ≤ 8, -8 ≤ l ≤ 9 |
| Reflections collected | 8054 |
| Independent reflections | 2628 [$R_{int}$ = 0.0261, $R_{sigma}$ = 0.0273] |
| Data/restraints/parameters | 2628/0/182 |
| Goodness-of-fit on F$^2$ | 1.075 |
| Final R indexes [I>=2σ (I)] | $R_1$ = 0.0412, $wR_2$ = 0.1150 |
| Final R indexes [all data] | $R_1$ = 0.0468, $wR_2$ = 0.1200 |
| Largest diff. peak/hole / e Å$^{-3}$ | 0.14/-0.14 |

To confirm the molecular structure, we prepared a large-sized crystal suitable for single-crystal X-ray diffraction (XRD) by the physical vapor transport (PVT) method. The BPDBNA hexagon microplates is monoclinic crystal system, belongs to the space group C2/c (CCDC No. 2324127) with cell parameters of *a* = 54.149 Å, *b* = 6.714 Å, *c* = 7.276 Å, and *α* = *γ* = 90°, *β* = 92.496°.

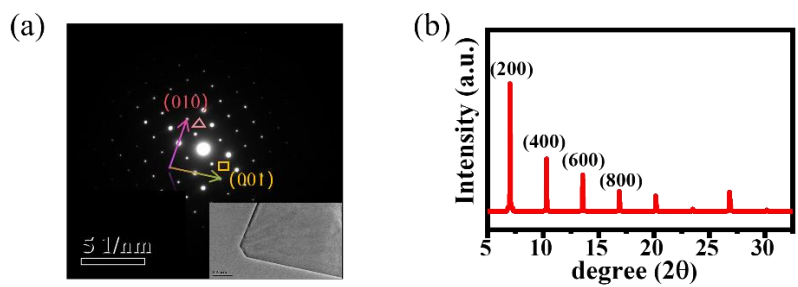

**Figure S6.** (a) Selected area electron diffraction (SAED) pattern of an individual BPDBNA single crystal and its corresponding transmission electron microscopy (TEM) image. The green circled and yellow triangled sets of spots in SAED pattern of BPDBNA crystals are ascribed to (010) and (001) Bragg reflections with d-spacing values of ~6.71 Å and ~7.20 Å, respectively. (b) XRD curve of BPDBNA crystals. By calculating the facet spacing, a series of peaks in the XRD spectra of the single crystal corresponds to the facets of the crystal. This suggests that the BPDBNA microcrystals adopt a lamellar structure in which the crystal (100) planes are parallel to the substrate.

XRD and selected area electron diffraction (SAED) results indicate that these hexagonal microplates are indeed single crystalline in nature, that grow preferentially along the (k00) crystal orientation.

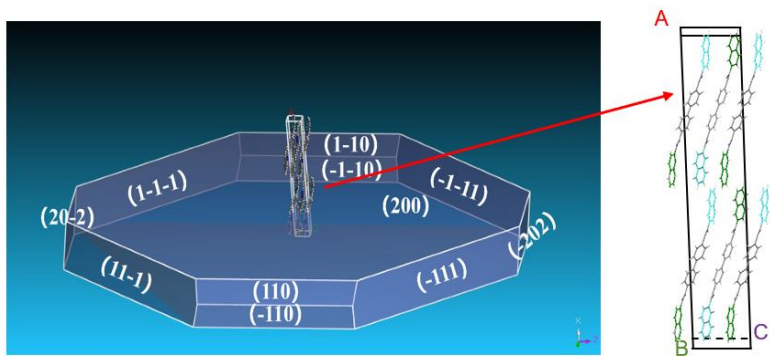

**Figure S7.** The simulated growth morphologies of BPDBNA crystal using Materials Studio package.

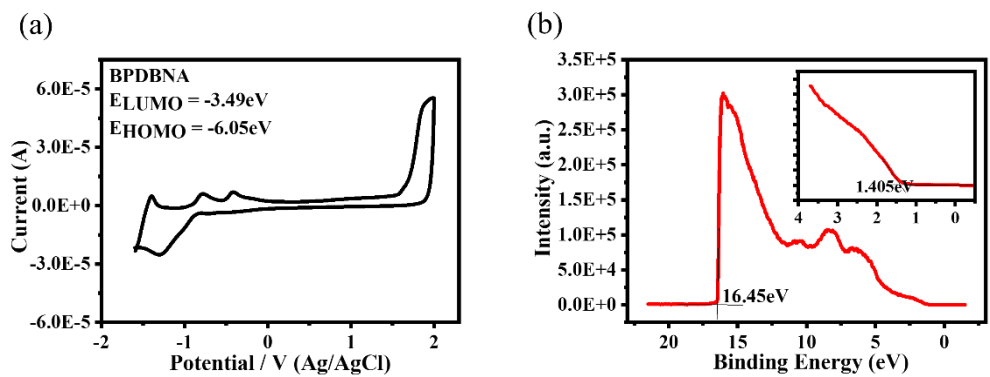

**Figure S8.** (a) Cyclic voltammetry (CV) curve of BPDBNA. (b) UPS spectrum of BPDBNA on ITO with an excitation energy of 21.22 eV. The onset of the ionization energy was determined to be 6.1 eV.

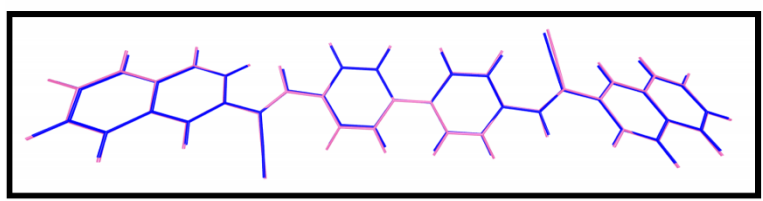

**Figure S9.** Comparison of theoretically calculated molecular configurations (blue) with experimentally obtained single-crystal structures (pink), RMSD is 0.15 Å and the difference between calculated and experimental emission energy is -0.12 eV.

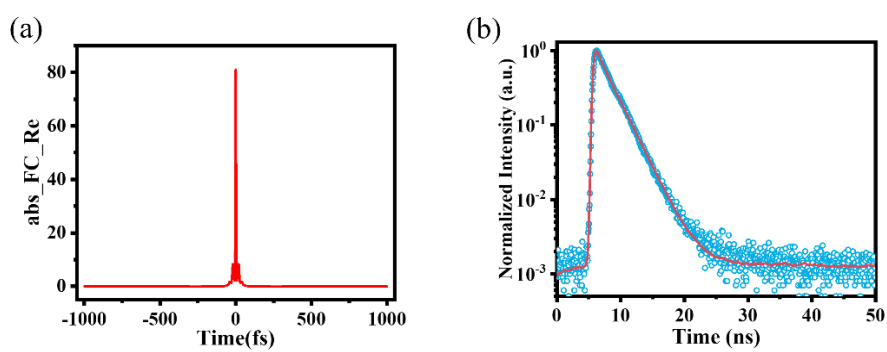

**Figure S10.** (a) Relevant convergence functions for theoretical calculations. (b) PL decay at 482 nm

**TableS2.** Comparison of Radiative Leap Rates from Theoretical Calculations and Experimental Data.

| Molecule | $\tau_{r\_cal}$ | $k_{r\_cal}$ | $\tau_{r\_exp}$ | $k_{r\_exp}$ |
|---|---|---|---|---|
| BPDBNA | 0.64 ns | $1.5653 \times 10^9$ s$^{-1}$ | 2.37 ns | $0.4219 \times 10^9$ s$^{-1}$ |

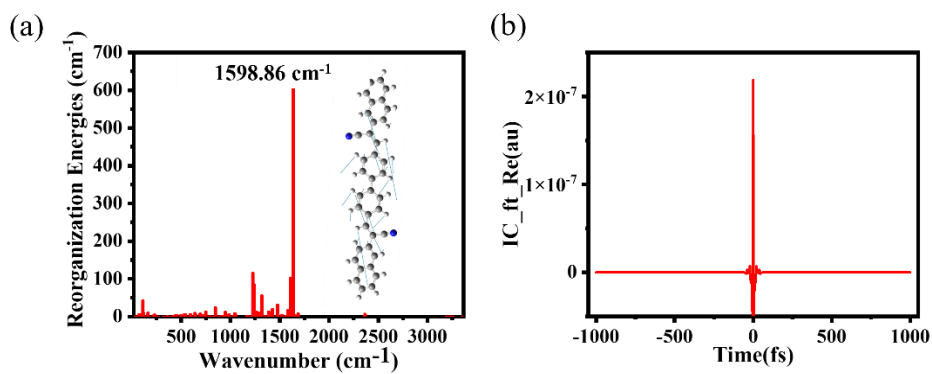

**Figure S11.** (a) Theoretical calculation of the reforming energy. Inset shows that the vibrations to which the reforming energy belongs are in-plane vibrations of hydrogen (H) atom in the skeleton. (b) Theoretical calculation of convergence functions associated with non-radiative jumps.

**TableS3.** Comparison of theoretically calculated radiative and non-radiative jump rates.

| Molecule | $\lambda_{S1-S0}$ | $k_{nr}$ | $k_r$ |
|---|---|---|---|
| BPDBNA | 0.1645 eV | $7.9289 \times 10^4$ s$^{-1}$ | $1.5653 \times 10^9$ s$^{-1}$ |

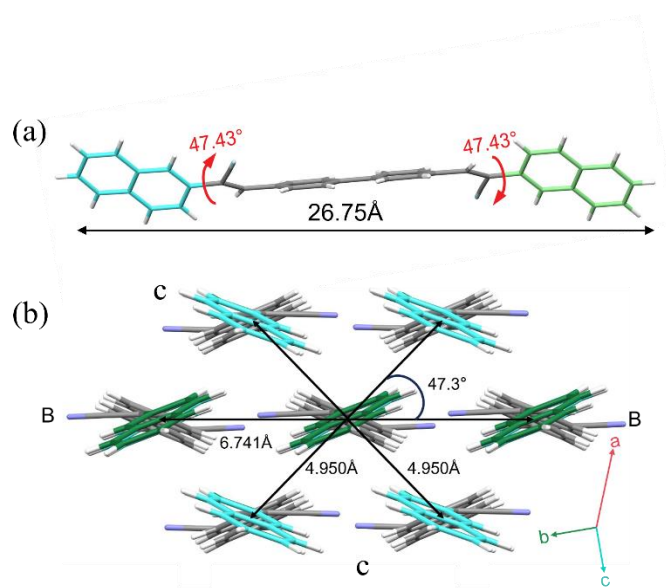

**Figure S12.** (a) Chemical structures of the BPDBNA molecule. And slight torsion angles exist between Two naphthalene ring groups (47.43°) relative to bi(phenyl-vinyl) skeleton, indicating that the rotational freedom provides conformation adjustment to balance optoelectronic properties. (b) Herringbone molecular packing with the *bc* plane.

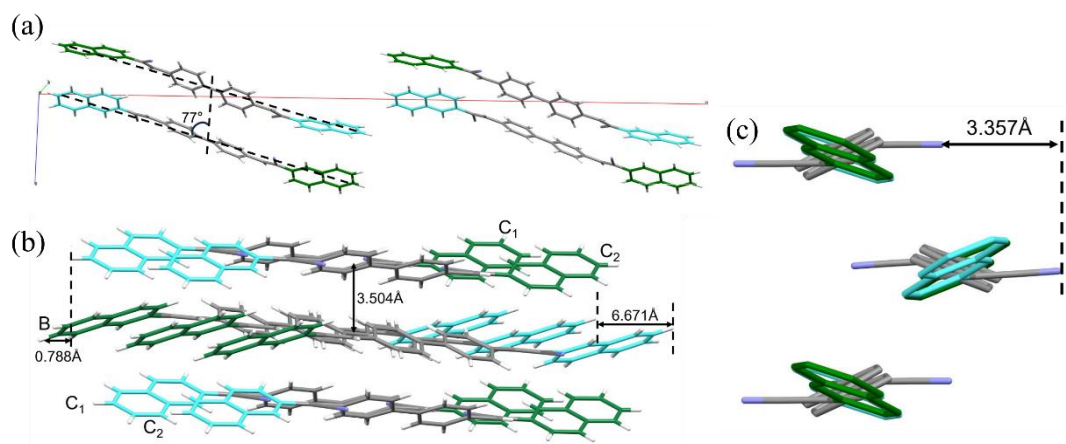

**Figure S13.** (a) The stacking angle along each π-column is determined to be 77° (>54.7°, that is the angle between the direction of π stacking and the direction of the transition dipole moment. (b) The longaxis displacements are relatively small ca. 0.788 and 6.671 Å for B and C pairs. (c) The BPDBNA molecule is complete stagger stacking along each π-column leading to a short-axis displacement as large as ~3.357 Å, which indicates a negligible π-π overlap and therefore weak π-π interactions.

To understand the efficient luminescence mechanism, we analyzed the molecular arrangement in the crystals and theoretically calculated the relevant photophysical parameters. According to the single-crystal cell-unit structure, BPDBNA molecules stack as a lamellar arrangement along the *a*-axis (Figure S6), with the layer height $a$ = 54.149 Å corresponding to the molecule length of 26.756 Å. Here, both naphthalene rings of BPDBNA molecule present symmetric torsion angle of 47.43° relative to bis(phenyl-vinyl) skeleton (Figure S9a). Notably, BPDBNA molecule adopt herringbone packing within the *bc* layer. Figure S9b depicts that each BPDBNA molecule was surrounded by six neighboring molecules, including two face-to-face π-π interacting molecules with a centroid-centroid distance of $d_{c-c}$ = 6.74 Å (labeled as B), fur edge-to-face CH-π interacting molecules with $d_{c-c}$ = 4.95 Å (labeled as C), and the herringbone angle between adjacent π-columns with 47.30°. The stacking angle along each π-column is determined to be 77° (Figure S10), which suggests that BPDBNA microcrystals are H-type aggregation according to Kasha's exciton mode[11], We calculated the excitonic couplings (J) by considering the coulomb dipole-dipole interactions at CAM-B3LYP/6-31G(d) level via NWchem6.3 packages[12,13]. Indeed, J > 0 was obtained for both face-to-face A pair as well as CH-π B cases (Table S2), clarifying H-type coupling[14]. In such H-type aggregates, the long-axis displacements are relatively small ca. 0.788 and 6.671 Å for B and C pairs (Figure S10), while the short-axis displacement along each π-column is as large as 3.357 Å, indicating a negligible π-π overlap and therefore weak π-π interactions. This endows strong emission and balanced charge transport of BPDBNA microplates.

**Table S4.** Computational Results of Transfer Integral V (Electronic Coupling), Excitonic Coupling J, Dispersion Energies $E_{disp}$, and Total Interaction Energies $E_{total}$ in Different Channels as Well as Reorganization Energy and Calculated Mobility.

| channel | Mobilities (cm$^2$V$^{-1}$s$^{-1}$) | J (meV) | $E_{disp}$ (kcal mol$^{-1}$) | $E_{total}$ (kcal mol$^{-1}$) | $E_{reorg,anion}$ (meV) | $E_{reorg,cation}$ (meV) | $V_h$ (meV) | $V_e$ (meV) |
|---|---|---|---|---|---|---|---|---|
| B | $\mu_h$=0.0120 | 60.467 | 43.842 | 3.897 | 160.36 | 98.337 | -7.751 | 1.637 |
| C | $\mu_e$=0.0297 | 79.295 | 51.972 | 0.989 | | | 26.025 | -18.219 |
| a1 | | -22.558 | 38.958 | 0.109 | | | -0.124 | 0.232 |
| a2 | | -19.685 | 38.647 | -0.099 | | | -0.235 | 0.77 |

To quantitatively describing intermolecular interactions, we calculated both dispersion energies $E_{disp}$ and total interaction energies $E_{total}$. Table 1 summarizes the results of A, B pairs as compared with the pair along *a* axis. It can be seen that intermolecular interactions in *bc* plane are much larger than those along *a* axis. Furthermore, the attachment energy $E_{attach}$ of the (001) crystal plane is much smaller than those within the ab plane, leading to the growth morphology of polygon microplate as predicted by the Materials Studio package (shown in Figure S6).

The sum values of $V_h$ and $V_e$ of BPDBNA are both comparably large; meanwhile, the cation and anion $E_{reorg}$ is calculated to be rather small, which open the possibility for BPDBNA with balanced ambipolar transport. Indeed, the calculated hole and electron mobilities of BPDBNA is relatively balanced, i.e., $\mu_h$ = 0.012 cm$^2$ V$^{-1}$ s$^{-1}$ and $\mu_e$ = 0.030 cm$^2$ V$^{-1}$ s$^{-1}$.

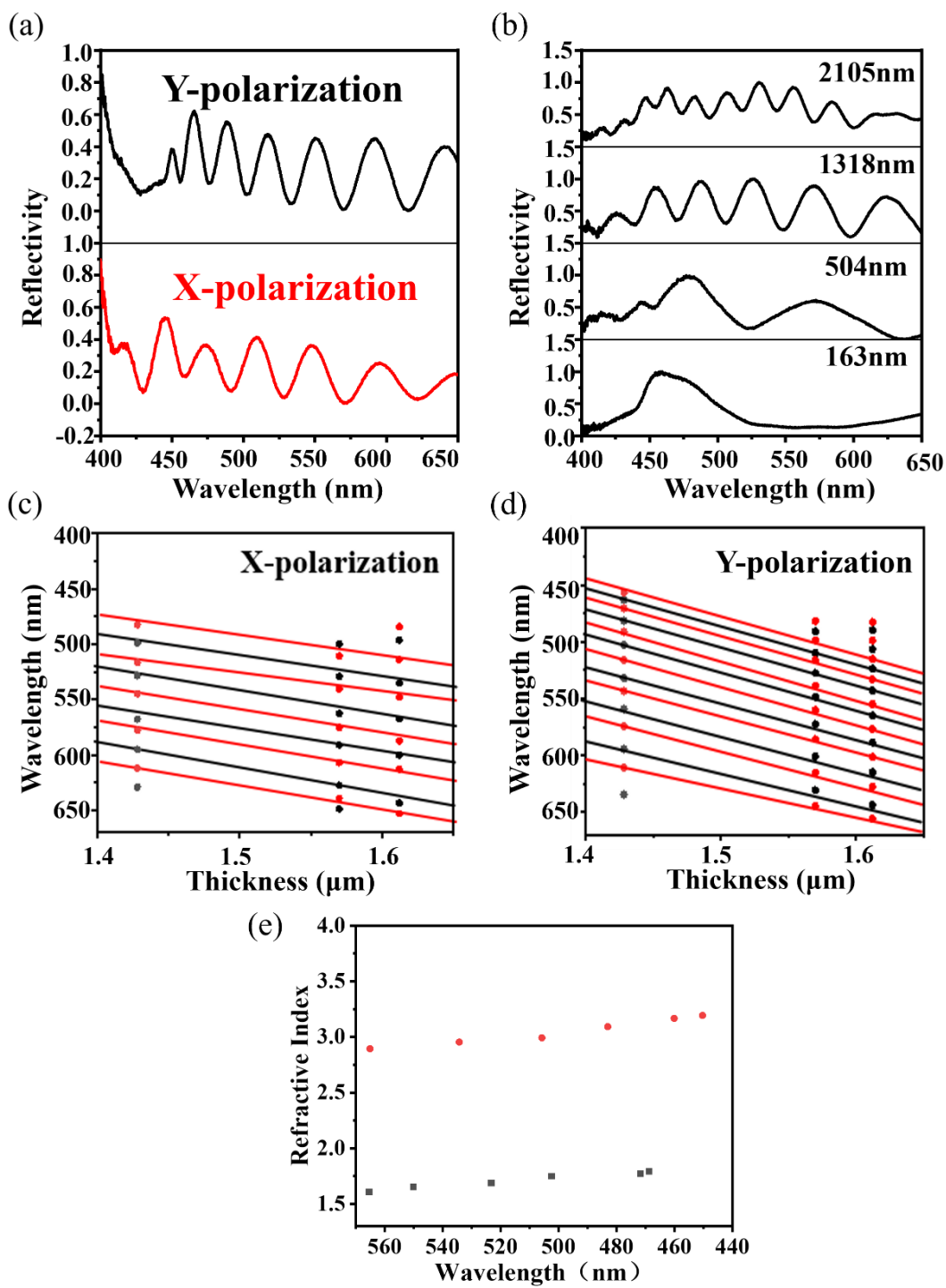

**Figure S14.** (a) Reflective spectra of the 1450-nm microcrystal cavity at the polarization of incident light parallel (black line) and vertical (red line) to the long direction of microcrystal. (b) Reflective spectra of the organic cavity with the thicknesses of 2105 nm, 1318 nm, 504 nm and 163 nm for parallel-polarization light. Wavelength of the interference maximum (red points) and minimum (black points) observed in reflection spectra of X-polarization (c) and Y-polarization (d). (e) The calculated refractive index of BPDBNA microcrystal in X-direction (black dot) and Y-direction (red dot). The n(λ) in the direction parallel to the long direction of the microcrystal rises from 2.89 at 570 nm to 3.19 at 450 nm, and rises from 1.60 at 570 nm to 1.79 at 450 nm in vertical direction. This is basically consistent with our simulation results.

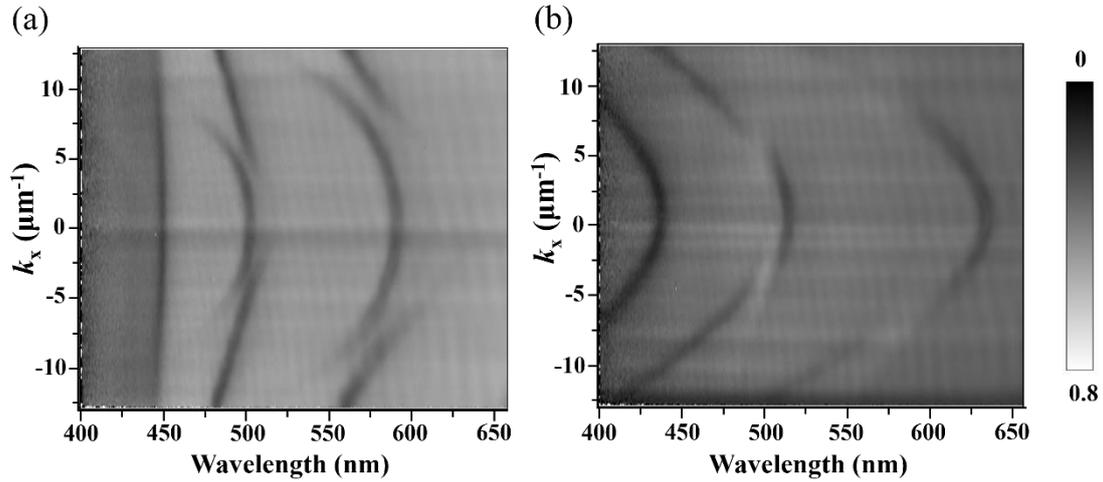

**Figure S15.** Measured angle-resolved reflectivity spectra of a selected microcavity at room temperature in X-polarization (a) and Y-polarization (b) along X-direction and Y-direction of the BPDBNA microcrystal, respectively.

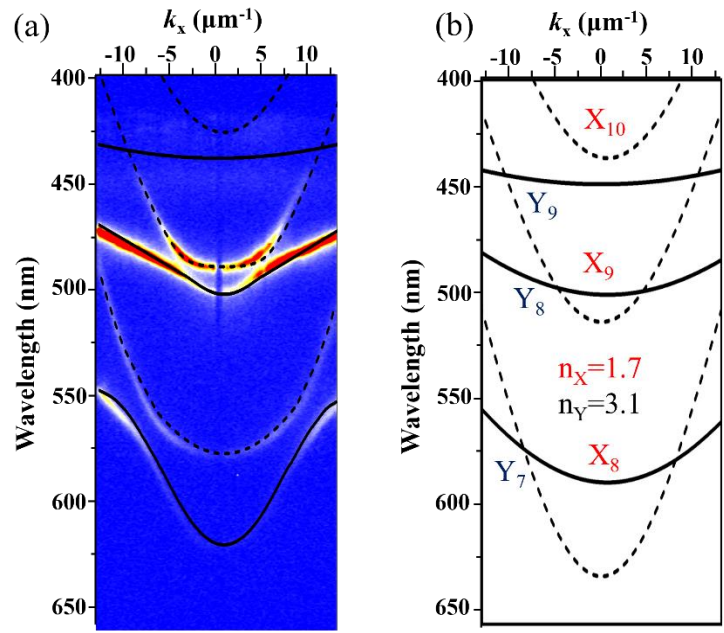

**Figure S16.** (a) The angle-resolved reflectivity spectra and the corresponding simulated cavity modes. (b) The calculated refractive indices are $n_x = 1.7$ and $n_y = 3.1$, respectively, corresponding to X-polarized mode and Y-polarized mode.

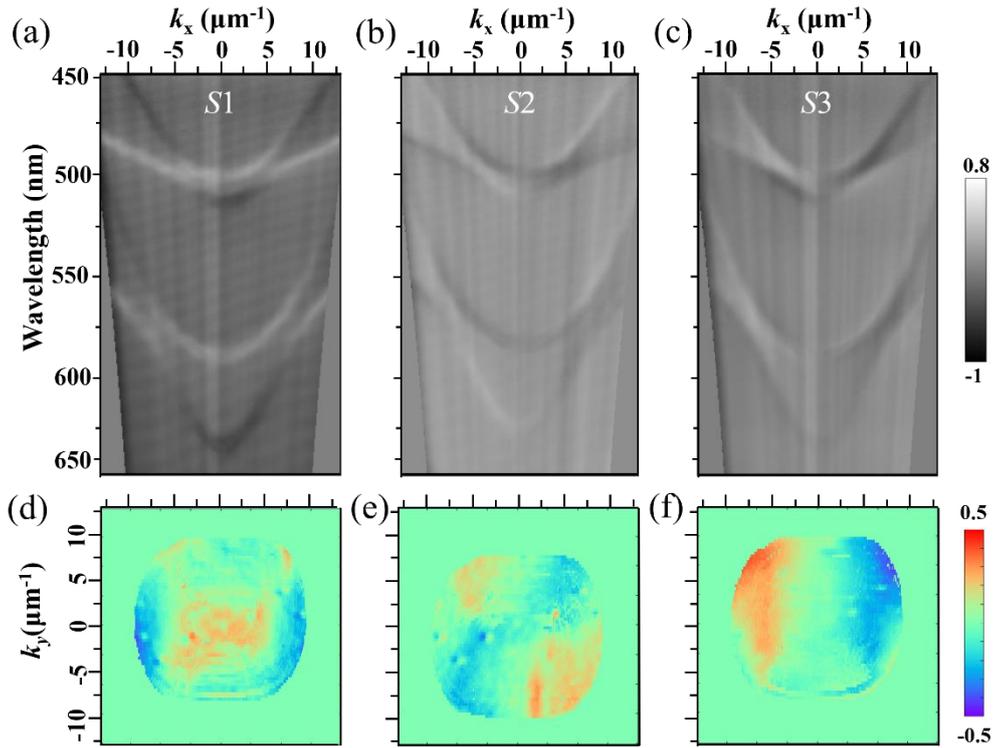

**Figure S17.** Angle-resolved reflectivity spectrum of the selected sample by employing a halogen lamp as the unpolarized white-light source. Measured Stokes parameters $S1$ (a), $S2$ (b) and $S3$ (c) of $X_9$, $Y_8$, $X_8$ and $Y_7$ modes. 2D maps of the Stokes parameters $S1$ (d), $S2$ (e) and $S3$ (f) of the X mode extracted from Figure 3a.

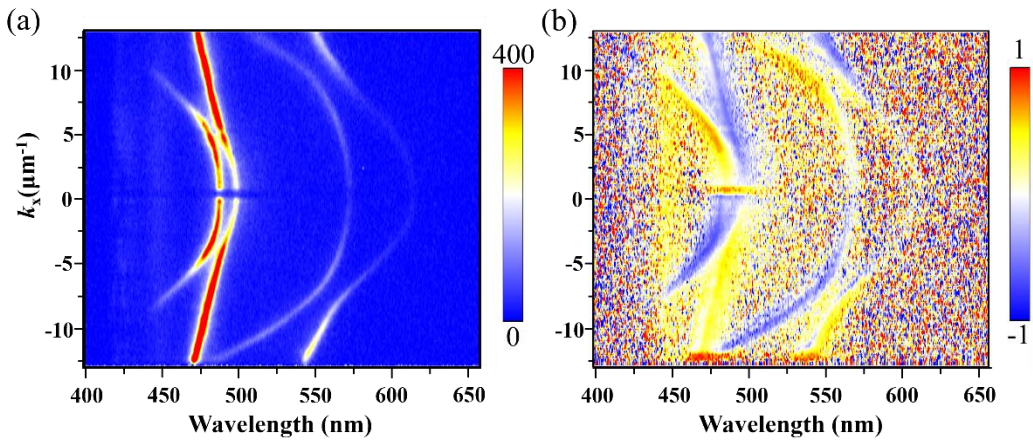

**Figure S18.** (a) The angle-resolved PL spectrum of the device. (b) S3 of the angle-resolved PL spectrum of the device.

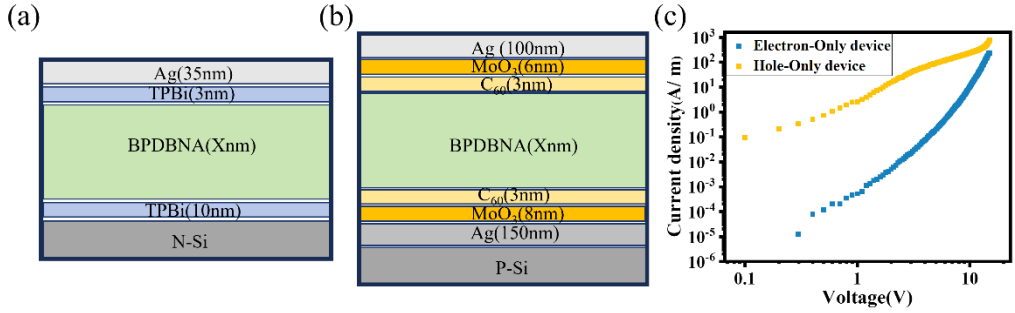

**Figure S19.** (a) and (b) BPDBNA electron-only and hole-only devices. (c) Current density-voltage characteristics of BPDBNA electron- (a) and hole-only (b) devices for 150nm thickness of the BPDBNA layer.

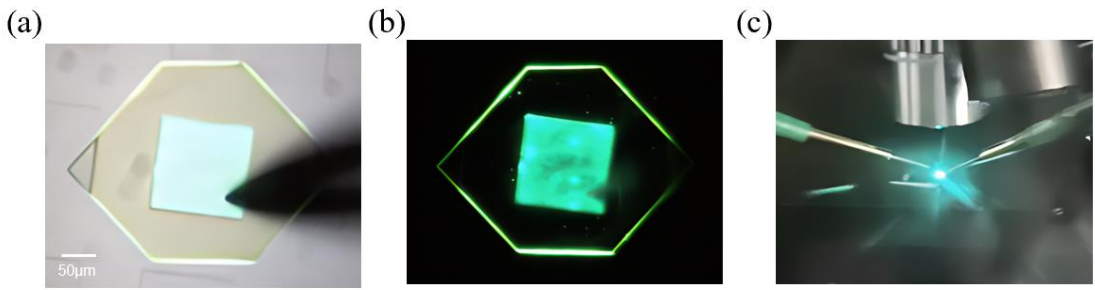

**Figure S20.** (a) Microscopic bright field image of BPDBNA crystal OLED. (b) Electroluminescent photographs of individual devices. Under electroluminescence, the device exhibits uniformly bright electroluminescence and edge waveguide. (c) Photographing the device after it has been illuminated.

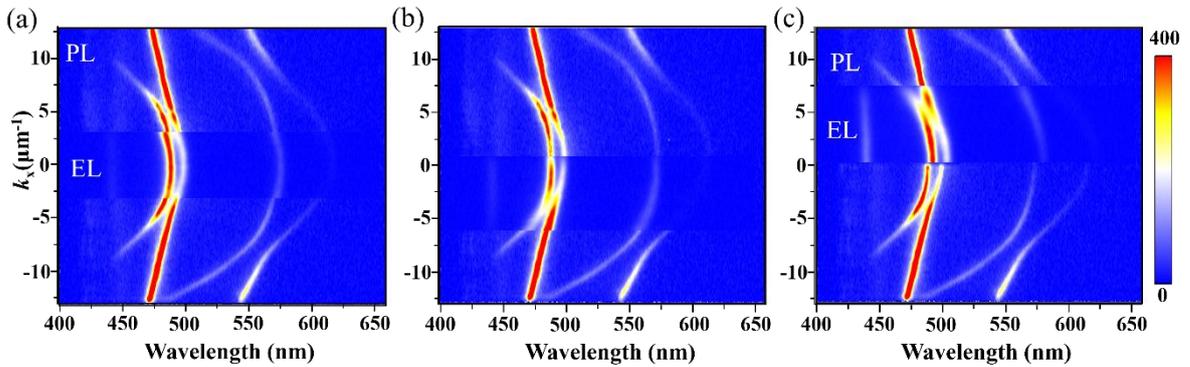

**Figure S21.** (a) Angle-resolved PL spectra under 405-nm laser excitation and angle-resolved EL spectra of the same device. (b) and (c) for angle-resolved EL spectra from other angles of the same device.

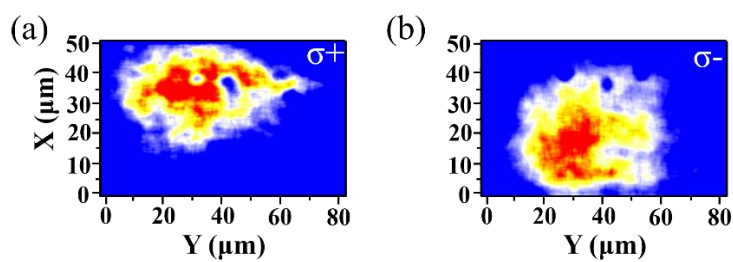

**Figure S22.** (a) and (b) Left- and Right-circularly polarized light in real space under electroluminescence.

**Reference:**

1. Lua, T. & Chen, F. Quantitative analysis of molecular surface based on improved Marching Tetrahedra algorithm. *J. Mol. Graph. Model.* **38**, 314-323 (2012).
2. Lu, T. & Chen, F. Multiwfn: A multifunctional wavefunction analyzer. *J. Comput. Chem.* **33**, 580-592 (2011).
3. Yin, F. *et al.* High-performance organic laser semiconductor enabling efficient light-emitting transistors and low-threshold microcavity lasers. *Nano Lett.* **22**, 5803-5809 (2022).
4. Marcus, R. A. Electron transfer reactions in chemistry. Theory and experiment. *Rev. Mod. Phys.* **65**, 599-610 (1993).
5. Niu, Y. *et al.* MOlecular materials property prediction pckage (MOMAP) 1.0: a software package for predicting the luminescent properties and mobility of organic functional materials. *Mol. Phys.* **116**, 1078-1090 (2018).
6. Nan, G. J., Yang, X. D., Wang, L. J., Shuai, Z. G. & Zhao, Y. Nuclear tunneling effects of charge transport in rubrene, tetracene, and pentacene. *Phys. Rev. B* **79**, 115203 (2009).
7. Shuai, Z., Wang, D., Peng, Q. & Geng, H. Computational evaluation of optoelectronic properties for organic/carbon materials. *Acc. Chem. Res.* **47**, 3301-3309 (2014).
8. Shuai, Z., Geng, H., Xu, W., Liao, Y. & Andre, J.-M. From charge transport parameters to charge mobility in organic semiconductors through multiscale simulation. *Chem. Soc. Rev.* **43**, 2662-2679 (2014).
9. Hashimoto, S., Ohno, N. & Itoh, M. Exciton–Polariton Dispersion of Thin Anthracene Crystals in the Thickness Range of 3 to 0.1 μm. *Phy. Status. Solidi. (b)* **165**, 277-286 (1991).
10. Wan, Y. *et al.* Efficient Organic Light-Emitting Transistors Based on High-Quality Ambipolar Single Crystals. *ACS Appl. Mater. Interf.* **12**, 43976-43983 (2020).
11. Kasha, M., Rawls, H. R. & EL-Bayoumi, M. A. The exction model in molecular spectroscopy. *Pure Appl. Chem.* **11**, 371-392 (1965).
12. Li, W. *et al.* Theoretical investigations on the roles of intramolecular structure

|    | distortion versus irregular intermolecular pcking in optical spectra of 6T nanoparticles. *Chem. Mater.* **29**, 2513-2520 (2017). |
|----|---|
| 13 | Jiang, Y., Shuai, Z. & Liu, M. Roles of long-range hopping, quantum nuclear effect, and exciton delocalization in exciton transport in organic semiconductors: a multiscale study. *J. Phys. Chem. C* **122**, 18365-18375 (2018). |
| 14 | Spano, F. C. Excitons in conjugated oligomer aggregates, films, and crystals. *Ann. Rev. Phys. Chem.* **57**, 217-243 (2006). |